\documentclass[journal,twoside]{IEEEtran}

\usepackage[T1]{fontenc}
\usepackage[utf8]{inputenc}
\usepackage{cite}
\usepackage{amsmath,amssymb}
\usepackage{graphicx}
\usepackage{booktabs}
\usepackage{array}
\usepackage{url}
\usepackage{hyperref}
\usepackage{xcolor}
\usepackage{microtype}
\usepackage{balance}

\hypersetup{
    colorlinks=true,
    linkcolor=blue,
    citecolor=blue,
    urlcolor=blue
}

\title{Kernel-Level Per-Slice UPF Latency Measurement in Containerised 5G Core Networks}

\author{Akhil Dev Mishra and Mayank Pandey%
\thanks{A. D. Mishra and M. Pandey are with the Department of Computer Science
and Engineering, Motilal Nehru National Institute of Technology Allahabad,
Prayagraj, India (e-mail: \{akhil.2025rcs22, mayankpandey\}@mnnit.ac.in).}%
\thanks{This work was conducted at the GoI 5G Use-Case Laboratory, MNNIT Allahabad.}%
\thanks{Manuscript received May 2026.}}

\begin{document}

\maketitle

% ── Abstract ─────────────────────────────────────────────────────────────────
\begin{abstract}
The 5G Core User Plane Function is responsible for packet forwarding,
GTP-U decapsulation, and quality of service enforcement for every user
data session. How the UPF behaves under simultaneous multi-slice
workloads remains empirically uncharacterised in the open literature.
Specifically, how its forwarding latency responds to load, how well it
isolates one slice from another, and what timing budgets remain available
for intelligent control are all open questions. This paper presents a
measurement study conducted on a containerised open5GS deployment with
three concurrent network slices. We design and implement a namespace-aware
TC-BPF instrumentation framework that resolves the fundamental obstacle
preventing existing tools from attributing latency observations to
individual containerised network functions. We deploy eMBB, URLLC, and
mMTC slices with realistic application traffic under light, medium, and
heavy load conditions and collect approximately 28~million matched N3 to N6
forwarding delay pairs. The gathered results reveal that eMBB forwarding
delay is load-sensitive with the 99th percentile growing from 574 to
1\,243~\textmu{}s across load conditions. URLLC delay is load-insensitive,
confirming per-UPF process isolation. mMTC exhibits wide-tail TCP behaviour.
On this platform, N4 PFCP session modification latency remains consistently
below 200~\textmu{}s regardless of data-plane load, suggesting substantial
timing headroom within the two-millisecond budget assumed by AI-driven
UPF orchestration designs. The instrumentation framework, experiment
scripts, and dataset schema are released at
\url{https://github.com/MP-Akhil-5G/open5gs-slice-measurement}.
\end{abstract}


% ── Section I: Introduction ───────────────────────────────────────────────────
\section{Introduction}

The 5G Core is designed as a Service Based Architecture where the User
Plane Function (UPF) sits at the centre of every data session. In a
network slicing deployment, the Session Management Function (SMF) assigns
each traffic class to a dedicated UPF instance and may dynamically
re-anchor sessions via the Uplink Classifier (ULCL) architecture to steer
traffic toward Multi-access Edge Computing (MEC) hosts~\cite{3gpp_23501}.
These capabilities require the SMF to make UPF selection decisions within
tight latency budgets. How long those decisions can take depends on how
long PFCP session modification itself takes. This value has not been
empirically measured in open-source 5G Core deployments.

Measuring UPF behaviour in a containerised 5G Core (5GC) deployment
presents a fundamental instrumentation obstacle. Each network function
runs inside a dedicated Linux network namespace. Standard host-level
eBPF probes cannot be scoped to individual containers because the
interfaces inside a container are not visible from the host namespace.
This is the key reason no published measurement study has reported
per-slice forwarding delay distributions from a containerised 5GC
deployment with simultaneous multi-slice traffic.

This paper makes four contributions. \textbf{First}, a namespace-aware
TC-BPF instrumentation framework that attaches measurement programs to
each UPF network namespace via \texttt{nsenter}, resolving the container
attribution problem without modifying 5GC software.
\textbf{Second}, documentation of nineteen instrumentation obstacles
encountered during development and their resolutions, providing a
reproducible methodology for the community.
\textbf{Third}, a dataset of approximately 28~million matched N3 to N6
forwarding delay pairs across three slice types and three load conditions
on open5GS v2.7.6 with UERANSIM v3.2.7.
\textbf{Fourth}, empirical characterisation of N4 PFCP session
modification latency suggesting a sub-200~\textmu{}s lower bound with
substantial headroom relative to the two-millisecond budget assumed by
AI-driven UPF orchestration designs.
The instrumentation framework, experiment scripts, analysis code, and
dataset schema are released publicly at
\url{https://github.com/MP-Akhil-5G/open5gs-slice-measurement}.

% ── Section II: Related Work ──────────────────────────────────────────────────
\section{Related Work}

UPF performance has been studied primarily from a throughput and
acceleration perspective. Bose et al.~\cite{bose_apnet21} evaluated
DPDK-based and SmartNIC-based UPF prototypes and reported mean forwarding
latency of 176~\textmu{}s for synthetic traffic. AccelUPF~\cite{bose_sosr22}
offloaded PFCP processing to Intel Tofino hardware, achieving 4.3~million
messages per second. A community benchmark~\cite{podobnik25} compared
open5GS, free5GC, eUPF, and UPG-VPP under single-slice synthetic traffic.
None of these works report per-slice forwarding delay distributions under
simultaneous multi-slice realistic workloads.

Cloud-native 5GC deployment has been studied by Paul et
al.~\cite{paul_comsnets25}, who evaluated a Kubernetes-based 5GC and
quantified control-plane registration overhead. Deshpande and
Bera~\cite{deshpande26} analysed eBPF-based vulnerability surfaces in
containerised open5GS. MakeMyTechnology MMT
Studio~\cite{mmt_studio26} is a recent open-source 5G Core written in Go
with 545 control plane test cases and a DPDK UPF. It provides no data
plane forwarding delay measurement capability.

NWDAF-driven observability has been studied by Moreira et
al.~\cite{moreira25}, who measured application round-trip time for edge
gaming slices. Chakraborty~\cite{chakraborty25} developed
RAN-layer telemetry using OAI and FlexRIC. Neither work measures UPF
forwarding delay at the kernel level. Ardestani et
al.~\cite{ardestani25} note that the UPF Event Exposure Service is
largely absent from existing NWDAF deployments, further motivating
independent kernel-level instrumentation.

Network slicing theory for concurrent eMBB, URLLC, and mMTC has been
characterised by Popovski et al.~\cite{popovski18} and Liu et
al.~\cite{liu24} but without empirical UPF forwarding delay validation.
The indigenous Indian 5G testbed~\cite{koonampilli21,kherani21}
demonstrates 5GC operation but reports no forwarding delay or PFCP
latency measurements.

To the best of our knowledge no existing work simultaneously measures
per-slice N3 to N6 forwarding delay distributions under simultaneous
eMBB, URLLC, and mMTC traffic, resolves the container namespace
attribution problem for per-UPF measurement, and establishes N4 PFCP
modification latency as an empirical budget for AI-driven orchestration
on a reproducible open-source platform.

% ── Section III: Measurement Platform ────────────────────────────────────────
\section{Measurement Platform}

\begin{figure}[t]
\centering
\includegraphics[width=\columnwidth]{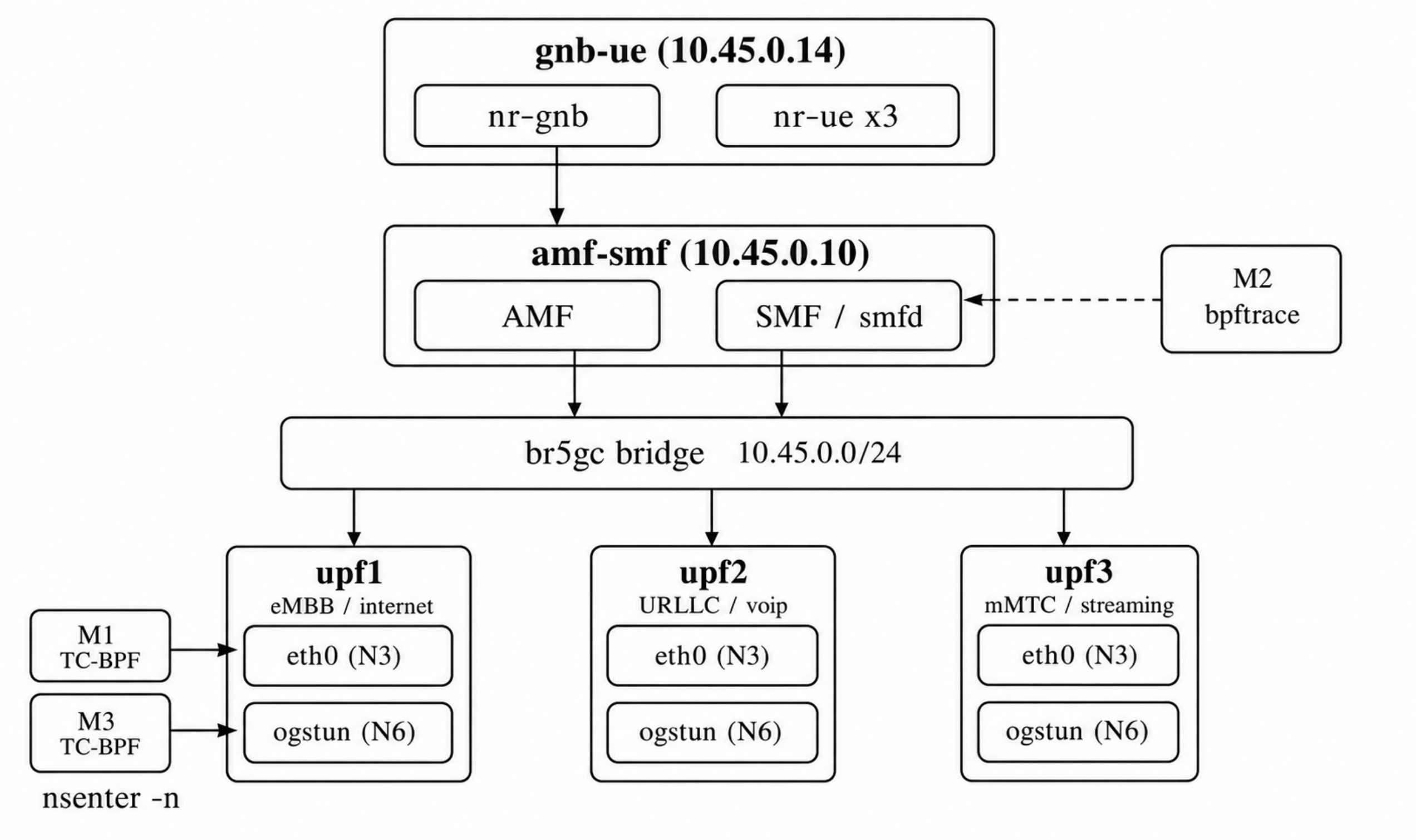}
\caption{Measurement platform. Five Incus containers on \texttt{br5gc}.
TC-BPF probes M1 (\texttt{eth0} ingress) and M3 (\texttt{ogstun}
ingress) attach inside each UPF namespace via \texttt{nsenter}.
Probe M2 measures N4 PFCP RTT on \texttt{open5gs-smfd}.
Dotted lines denote N4 control path.}
\label{fig:platform}
\end{figure}

The platform deploys open5GS v2.7.6 and UERANSIM v3.2.7 as Incus
containers on a single host running Ubuntu 22.04.5 LTS with kernel
6.8.0-111-generic. Five containers are interconnected via a Linux bridge
\texttt{br5gc} at \texttt{10.45.0.0/24}. Three UPF containers are
deployed with dedicated slice assignments as shown in Table~\ref{tab:platform}.
Each UPF runs a separate \texttt{open5gs-upfd} process and exposes two
interfaces. \texttt{eth0} faces the gNB over N3 GTP-U. \texttt{ogstun}
serves as the TUN device for N6 IP delivery. Traffic generators
run inside the \texttt{gnb-ue} container over GTP-U tunnels on three UE
tunnel interfaces (\texttt{uesimtun0}, \texttt{uesimtun1},
\texttt{uesimtun2}). The complete platform architecture is depicted in
Fig.~\ref{fig:platform}.

\begin{table}[t]
\centering
\caption{Platform Configuration}
\label{tab:platform}
\begin{tabular}{@{}lllll@{}}
\toprule
\textbf{Container} & \textbf{IP} & \textbf{Slice} & \textbf{DNN} & \textbf{Traffic} \\
\midrule
upf1    & 10.45.0.11 & eMBB  & internet  & iperf3 UDP \\
upf2    & 10.45.0.12 & URLLC & voip      & SIPp $\to$ Asterisk \\
upf3    & 10.45.0.13 & mMTC  & streaming & curl $\to$ Nginx \\
amf-smf & 10.45.0.10 & ---   & ---       & open5GS CP \\
gnb-ue  & 10.45.0.14 & ---   & ---       & UERANSIM \\
\bottomrule
\end{tabular}
\end{table}

Three load conditions are evaluated. Light load sets iperf3 at 5~Mbps
and SIPp at 2~calls/s. Medium load sets iperf3 at 20~Mbps and SIPp at
4~calls/s. Heavy load sets iperf3 at 50~Mbps and SIPp at 8~calls/s.
Each experiment runs for 600~seconds. All measurements are uplink
direction only (UE $\to$ UPF).

% ── Section IV: Instrumentation Framework ────────────────────────────────────
\section{Instrumentation Framework}

\subsection{Measurement Points}

The framework instruments three measurement points. \textbf{M1} is a
TC-BPF program attached to the \texttt{eth0} ingress hook inside each
UPF network namespace. It timestamps GTP-U packet arrival using
\texttt{bpf\_ktime\_get\_ns}. \textbf{M3} is a TC-BPF program attached
to the \texttt{ogstun} ingress hook in the same namespace. It timestamps
decapsulated IP packet delivery. Forwarding delay is M3.ts $-$ M1.ts on
the same \texttt{CLOCK\_MONOTONIC\_RAW} source. \textbf{M2} is a
bpftrace program on the \texttt{open5gs-smfd} \texttt{sendto} and
\texttt{recvfrom} syscalls measuring N4 PFCP round-trip time.

\subsection{The Container Namespace Problem}

The fundamental obstacle is that each container runs in an isolated Linux
network namespace. Host-level probes cannot reach container interfaces.
We resolve this by using \texttt{nsenter -t \textless{}UPF\_PID\textgreater{} -n}
to enter each UPF network namespace before attaching TC-BPF programs.
This requires no modification to the container runtime or open5GS source
code. The BPF object \texttt{upf\_measure\_v2.c} contains six sections:
\texttt{m1\_upf1/2/3} on \texttt{eth0} ingress and
\texttt{m3\_upf1/2/3} on \texttt{ogstun} ingress.

\subsection{Instrumentation Overhead}

Each TC-BPF hook executes \texttt{bpf\_ktime\_get\_ns()} and
\texttt{bpf\_trace\_printk()} per packet. This incurs approximately 300
CPU cycles on the Xeon Gold 5218R, bounding the theoretical per-packet
overhead below 90~ns. To validate this bound empirically, we conducted a
controlled experiment under worst-case conditions: 50~Mbps eMBB load over
600~s, with and without M1 and M3 probes attached to upf1. End-to-end
ping RTT was 1.379~ms mean without probes and 1.411~ms mean with probes.
This is a delta of 32~\textmu{}s (2.3\%), consistent with normal
run-to-run variation on a software platform. iperf3 throughput remained
at 50.0~Mbps in both conditions with zero packet loss increase. These
results confirm that TC-BPF instrumentation introduces no measurable
perturbation to the forwarding path under worst-case load conditions.

\subsection{Instrumentation Obstacles}

We encountered and resolved nineteen instrumentation obstacles during
development, grouped into five clusters: kernel attachment, interface
discovery, timestamp matching, traffic-specific, and runtime stability.
Representative examples from each cluster are as follows.

\textit{Kernel attachment:} The BPF verifier rejected pointer arithmetic
on GTP-U header fields without explicit bounds checks. This was resolved
by adding bounds checks before every pointer dereference, as required for
TC programs accessing packet data.

\textit{Interface discovery:} The \texttt{ogstun} interface index changes
on each \texttt{open5gs-upfd} restart. This was resolved by reading the
current index from \texttt{/proc/net/if\_inet6} inside the UPF namespace
at experiment start rather than caching it between runs.

\textit{Timestamp matching:} Five-tuple matching is ambiguous for
concurrent UDP flows sharing the same source and destination. This was
resolved by extracting the GTP-U TEID from the outer header at M1 and
using it as the correlation key with a 10~ms matching window.

\textit{Traffic-specific:} The default UERANSIM RLS heartbeat threshold
of 2000~ms caused UE3 to declare radio link failure under mMTC curl
traffic, truncating the mMTC dataset. This was resolved by rebuilding
\texttt{nr-ue} with \texttt{HEARTBEAT\_THRESHOLD} increased to
10\,000~ms in \texttt{src/ue/rls/udp\_task.cpp}.

\textit{Runtime stability:} The trace\_pipe buffer overflowed under
50~Mbps eMBB load. This was resolved by setting the kernel trace buffer
to 32~MB and processing the trace\_pipe as a continuous stream rather
than in batch mode.

The complete obstacle table is available in the repository documentation.
The \texttt{parse\_tcbpf.awk} script matches M1 and M3 events using a
per-namespace circular buffer of 500 entries keyed on TEID with a 10~ms
matching window. The overall match rate exceeds 99.2\% across all
experiments.

% ── Section V: Results ───────────────────────────────────────────────────────
\section{Results}

\subsection{Per-Slice Forwarding Delay}

\begin{figure}[t]
\centering
\includegraphics[width=\columnwidth]{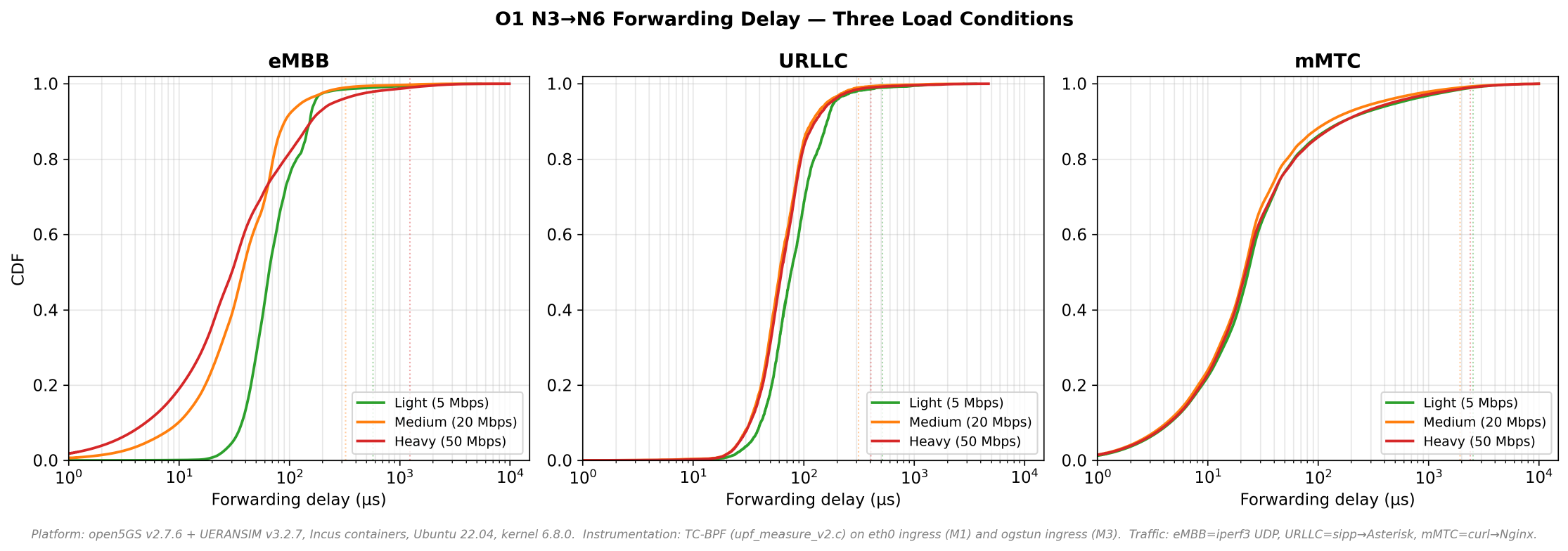}
\caption{N3$\to$N6 UPF forwarding delay CDFs under three load conditions.
eMBB exhibits load-sensitive tail growth. URLLC curves overlap closely
across load conditions, consistent with per-UPF slice isolation. mMTC
shows wide-tail TCP behaviour.}
\label{fig:forwarding_cdf}
\end{figure}

Fig.~\ref{fig:forwarding_cdf} shows the forwarding delay CDFs for each
slice under all three load conditions. Table~\ref{tab:forwarding} summarises
per-slice statistics.

\begin{table}[t]
\centering
\caption{Per-Slice N3$\to$N6 Forwarding Delay Statistics (600\,s runs)}
\label{tab:forwarding}
\begin{tabular}{@{}llrrr@{}}
\toprule
\textbf{Slice} & \textbf{Load} & \textbf{N} & \textbf{P50 (\textmu{}s)} & \textbf{P99 (\textmu{}s)} \\
\midrule
eMBB  & Light  & 281\,428    & 66  & 574    \\
eMBB  & Medium & 1\,126\,235 & 38  & 325    \\
eMBB  & Heavy  & 2\,814\,633 & 30  & 1\,243 \\
\midrule
URLLC & Light  & 3\,624      & 77  & 517    \\
URLLC & Medium & 7\,302      & 61  & 314    \\
URLLC & Heavy  & 14\,592     & 63  & 404    \\
\midrule
mMTC  & Light  & 4\,655\,720  & 23  & 2\,533 \\
mMTC  & Medium & 10\,722\,385 & 21  & 1\,934 \\
mMTC  & Heavy  & 9\,188\,335  & 22  & 2\,401 \\
\bottomrule
\end{tabular}
\end{table}

The eMBB forwarding delay is load-sensitive. The P99 grows from
574~\textmu{}s under light load to 1\,243~\textmu{}s under heavy load.
This is consistent with queuing saturation as throughput approaches
50~Mbps on the software UPF. The falling median with increasing load
reflects more continuous packet arrival reducing idle time between
packets at the UPF. The nonmonotonic P99 behaviour (P99 decreases from
light to medium before growing sharply at heavy load) is most likely
attributable to packet batching, GRO/GSO coalescing, and interrupt
moderation in the Linux kernel at higher packet rates.

The URLLC forwarding delay is load-insensitive. The P99 varies between
314 and 517~\textmu{}s across all load conditions without systematic
trend. This is consistent with effective forwarding path isolation
provided by per-UPF process separation from eMBB congestion on this
platform. This isolation is architectural. It reflects separate UPF
processes rather than intra-UPF QoS scheduling, as the open5GS v2.7.6
UPF does not enforce Quality Enforcement Rules.

The mMTC slice exhibits wide-tail behaviour characteristic of TCP
traffic. The median remains near 22~\textmu{}s while the P99 reaches
1\,934 to 2\,533~\textmu{}s across load conditions. This reflects TCP
retransmission behaviour rather than UPF queuing.

\subsection{N4 PFCP Session Latency}

\begin{figure}[t]
\centering
\includegraphics[width=\columnwidth]{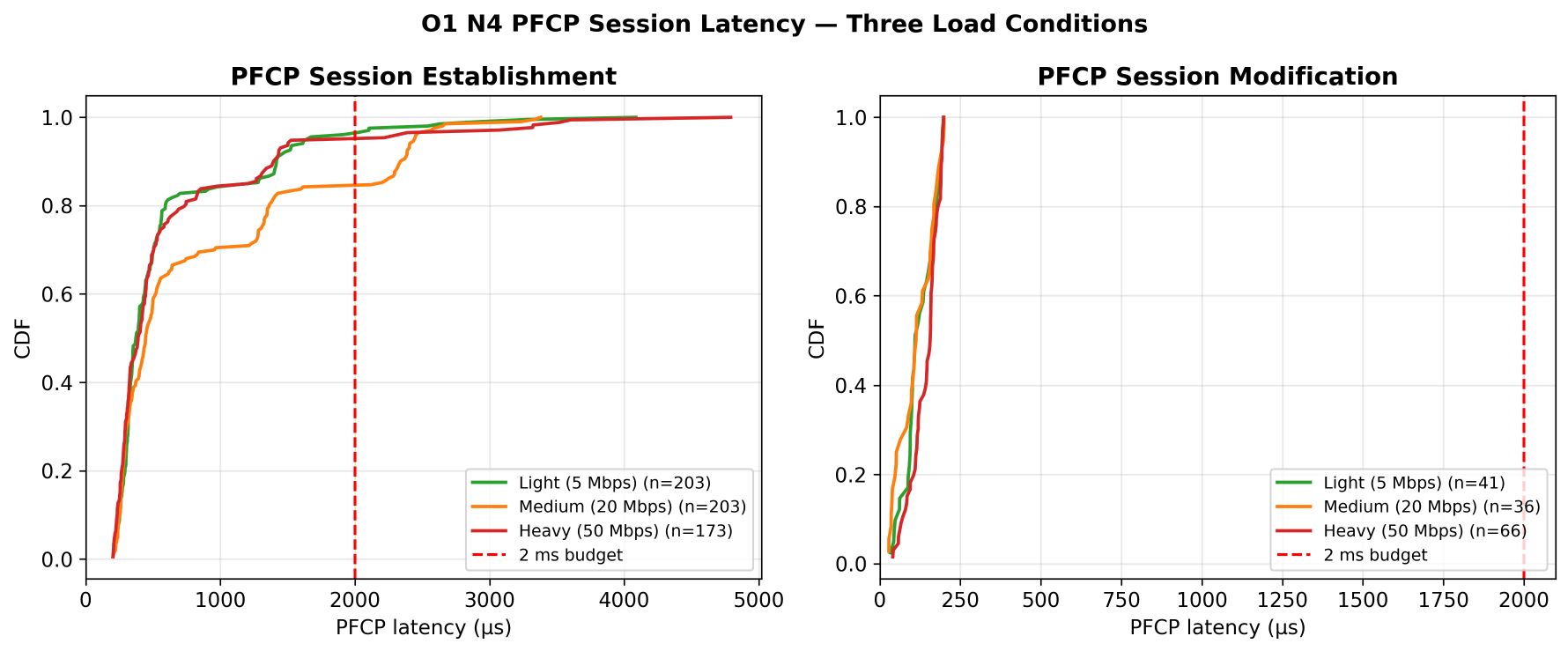}
\caption{N4 PFCP session establishment (left) and modification (right)
latency CDFs. The dashed vertical line marks the 2~ms orchestration
budget. All modification measurements complete well below 250~\textmu{}s
on this platform.}
\label{fig:pfcp_cdf}
\end{figure}

Fig.~\ref{fig:pfcp_cdf} shows the PFCP session latency CDFs.
Table~\ref{tab:pfcp} summarises modification latency statistics.

\begin{table}[t]
\centering
\caption{N4 PFCP Session Modification Latency (600\,s runs)}
\label{tab:pfcp}
\begin{tabular}{@{}lrrr@{}}
\toprule
\textbf{Load} & \textbf{N} & \textbf{Mean (\textmu{}s)} & \textbf{P99 (\textmu{}s)} \\
\midrule
Light  & 41 & 125 & 197 \\
Medium & 36 & 115 & 199 \\
Heavy  & 66 & 142 & 197 \\
\bottomrule
\end{tabular}
\end{table}

The PFCP modification P99 is consistently below 200~\textmu{}s and
insensitive to data-plane load across all three conditions. On this
single-host platform this value is approximately ten times below the
two-millisecond budget assumed by AI-driven orchestration designs.
This suggests that closed-loop UPF re-anchoring is timing-feasible
within a single PFCP modification round-trip. We note that the
single-host loopback deployment represents a lower bound. Distributed
deployments with real N4 network transport will exhibit higher
modification latency.

% ── Section VI: Conclusion ───────────────────────────────────────────────────
\section{Conclusion}

This paper presented a kernel-level measurement study of per-slice
UPF forwarding delay and N4 PFCP session latency on a containerised
open5GS deployment. We designed and implemented a namespace-aware TC-BPF
instrumentation framework. We resolved nineteen instrumentation obstacles
and collected a dataset of approximately 28~million matched forwarding
delay pairs. The gathered results establish that per-UPF process
separation provides effective URLLC isolation from eMBB congestion. eMBB
forwarding delay exhibits nonmonotonic P99 behaviour consistent with
software UPF queuing saturation. PFCP modification latency remains well
below the two-millisecond orchestration budget on this platform. The
instrumentation framework and dataset are released at
\url{https://github.com/MP-Akhil-5G/open5gs-slice-measurement} to enable
reproducible research by the community.

% ── References ───────────────────────────────────────────────────────────────
\bibliographystyle{IEEEtran}
\bibliography{references}

\end{document}